\documentstyle[aps,prl,twocolumn,graphicx]{revtex}

\begin{document}

\title{Memory beyond memory in heart beating: an efficient way to detect 
pathological conditions}
\author{P. Allegrini$^{1}$, P. Grigolini$^{2,3,4}$, P. Hamilton$^{5}$,
L. Palatella$^{3}$, G. Raffaelli$^{3}$}
\address{$^{1}$ Istituto di Linguistica Computazionale del Consiglio Nazionale delle
Ricerche,\\
 Area della Ricerca di Pisa-S. Cataldo, Via Moruzzi 1, 
56124, Ghezzano-Pisa, Italy }
\address{$^{2}$Center for Nonlinear Science, University of North Texas,
P.O. Box 311427, Denton, Texas, 76203-1427}
\address{$^{3}$Dipartimento di Fisica dell'Universit\`{a} di Pisa and INFM 
Piazza Torricelli 2, 56127 Pisa, Italy }
\address{$^{4}$Istituto di Biofisica del Consiglio Nazionale delle
Ricerche,\\ Area della Ricerca di Pisa-S. Cataldo, Via Moruzzi 1,
56124, Ghezzano-Pisa, Italy }
\address{$^{5}$ Center for Nonlinear Science, Texas Woman's University, 
P.O. Box 425498, Denton, Texas 76204}
\date{\today}
\maketitle

\begin{abstract}
 We study the long-range correlations of heartbeat fluctuations with 
the method of diffusion entropy. We show that this method of 
analysis yields a scaling
parameter $\delta$ that apparently conflicts with the direct evaluation 
of the distribution of times of sojourn in states with a given 
heartbeat frequency. The strength of the memory responsible for this 
discrepancy is given by a parameter $\epsilon^{2}$, which is derived 
from real data. The distribution of patients in the ($\delta$, 
$\epsilon^{2}$)-plane yields a neat separation of the healthy from the 
congestive heart failure subjects.
\vspace{.3 cm}

\end{abstract}
\vspace{.3 cm}

The analysis of time series of physiological 
significance is currently done using the paradigm 
of anomalous scaling\cite{shlesinger}. This letter, resting on 
this paradigm, aims at showing that the entropy of 
a diffusion process generated by a physiological 
time series according to the prescriptions of 
Refs.\cite{nicola,giacomo} 
yields a 
scaling exponent that depends only on genuine events, 
namely, events whose occurrence time is unpredictable.
We call this method of analysis Diffusion Entropy (DE) method and,
by means of simple dynamic models,
 we prove it to be insensitive to pseudo events, namely, to events whose 
occurrence times are correlated to those of earlier events.

\begin{figure}[!h]
\begin{center}
\includegraphics[scale=.31]{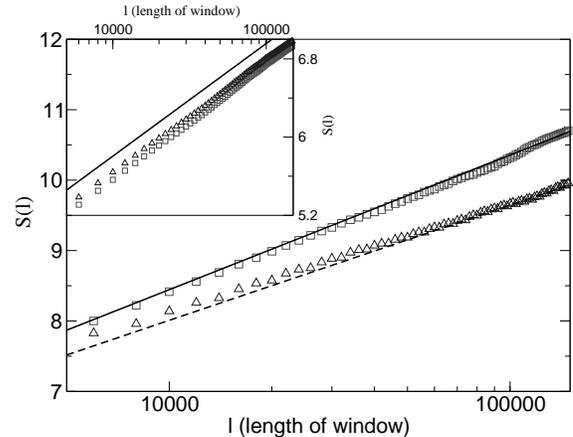}
\caption{\label{fig1}
DE for two-variables model as a function of time. The squares correspond to
$k'$=0.018, $z'$=1.83 while the diamonds to $k'$=0.011, $z'$=1.71. For both
curves $k$=0.4, $z$=1.25. In the inset: the same curves after shuffling,
the straight line slope is $0.5$.}
\end{center}
\end{figure}

         Let us consider first a dynamic model that generates events 
(really random events). This is given by
\begin{equation}
    \label{keyequation}
    \dot x  =  \Phi(x) > 0,
\end{equation}
where $x$ denotes the coordinate of a particle, moving within the 
interval $I \equiv [0,1]$, from the left to the right, with times of 
arrival at $x = 1$ determined by Eq.(\ref{keyequation}) and by the 
initial condition. 
When the particle reaches the right border of $I$, it is injected back 
to a new initial condition selected with uniform probability on $I$. 
Consequently,  the  times of arrival at $x = 1$, $t_{1},..,t_{i}..,$ 
are a fair example of real events.
It is straightforward to prove that the choice $\Phi(x) = \kappa x^{z}$,
with $z > 1$ and $k > 0$, 
yields for the waiting times $\tau_{i} \equiv t_{i}- t_{i-1}$ the 
following distribution density
\begin{equation}\label{equation2}
    \psi(\tau) = (\mu -1) \frac{T^{\mu -1}}{(T + \tau)^{\mu}},
    \end{equation}
    with $\mu = z/(z-1)$ and 	$T = (\mu-1)/\kappa$. 
 Note that the mean  waiting time 
$ \langle \tau \rangle$ is determined by $T$ through 
$\langle \tau \rangle = T/(\mu - 2)$.    
Let us convert now the time series $\{\tau_{i}\}$ into a random walk.
We select the rule\cite{giacomo} that makes the random walker move, 
always in the same 
direction and by a step of constant intensity, only when an event occurs. 
This means that the sequence $\{\tau_{i}\}$ is converted into a sequence 
of 0's and 1's as follows. We create a sequence of patches, one patch 
for each $\tau_{i}$, with a width given by the integer part of 
$\tau_{i}$; we fill the sites of each patch with $0$'s and we signal the 
border between two nearest-neighbor patches with $1$'s.
Then, the resulting sequence 
is converted into many trajectories of a given length $l$ with 
the method of the moving window. A window of size $l$ moves along the 
sequence and for any window position, the  portion of the whole sequence 
spanned by the window is regarded as a single trajectory of length $l$. 
All these 
trajectories are assumed to start from the origin, and are used to 
create a diffusion distribution, at time $l$. If there is scaling, 
with scaling parameter $\delta$, the 
entropy $S(l)$ takes the form\cite{nicola,giacomo}
\begin{equation}
    S(l) = A + \delta \ln (l) .
    \label{entropyform}
    \end{equation}
    Consequently, the rate of entropy increase in logarithmic 
    time scale is the 
value of the scaling parameter we are looking for. 
     The DE method has been widely discussed in earlier 
    papers\cite{giacomo,vito} and it has been pointed out that, without 
    using any form of detrending, it yields the correct scaling 
    parameter $\delta$. 
In some cases\cite{vito} this method detects the 
L\'{e}vy scaling that  would imply an 
infinite variance and, consequently, would be incompatible 
    with the adoption of methods based on variance. 
In the specific case of the model of 
    Eq.(\ref{keyequation}) the adoption of this method of analysis yields a 
    scaling parameter $\delta$ that, in the case where $2< \mu < 3$, fits the 
    theoretical prediction
    \begin{equation}
	\label{nomemory}
	\delta = \frac{1}{\mu -1},
	\end{equation}
	while for $\mu > 3$ one gets $\delta = 0.5$.
	This is a proof that the direct evaluation of the power law index 
	$\mu$ is equivalent to detecting the scaling $\delta$ by means of the 
	DE method.  
    It is well known \cite{katia} that a Markov master equation, 
    namely a stochastic process without memory, is characterized by a 
    waiting time distribution $\psi(\tau)$ with an exponential form, 
    thereby implying that a marked deviation from the 
    exponential condition is a signature of the presence of memory. 
      This is the kind of memory that is usually associated with the 
    detection of an anomalous scaling parameter, namely $\delta \neq 0.5$. 
    We refer ourselves to this memory as memory of the first type.
      To prepare the ground for the kind of memory that is 
	the focus of this paper, we have to discuss a dynamic model 
	generating both events and pseudo events.
For this purpose let us consider a two-variables model. The first 
equation, referring to the first variable, is given by Eq.(\ref{keyequation}), 
and the second equation,
concerning the new variable $y$, is given by
    \begin{equation}
    \label{keyequation2}
    \dot y  =  \chi(y) > 0.
   \end{equation}
The variables $x$ and $y$ are the coordinates of two particles, both moving
in the interval $I$, always from the left to the right. The initial 
conditions of the variable $y$ are always chosen randomly. The initial 
conditions of $x$, on the contrary, are not always chosen randomly, but 
they are only when the variable $y$ reaches the border at least once, 
during the sojourn of $x$ within the interval. Let us consider the 
sojourn time interval $[t_{i},t_{i+1}]$. If in this time interval the 
variable y remains within the interval, without touching the right 
border, then we set $x(t_{i+1}) = x(t_{i})$. This is a pseudo event. 
Thus, the sequence $\{t_{i}\}$ is a mixture of events and pseudo events. 
Let us consider the case where $\chi(y) = k'y^{z'}$ with $z' > 1$
and $k' > 0$, so as to produce 
the power index $\mu'=z'/(z'-1)$, with $\mu' > 2$, a property of real 
events. Let us set the condition 
$\langle \tau \rangle_{x} \ll \langle \tau \rangle_{y}$. In this case 
it is straightforward to show that the waiting time distribution of $x$ 
is still given by Eq.(\ref{equation2}). However, the power index $\mu$ 
is not more a reflection of real events.
Fig.\ref{fig1} reveals a very attractive property: the DE now yields 
$\delta = 1/(\mu'-1)$, which is very different from the theoretical prediction 
of Eq.(\ref{nomemory}) that makes $\mu > 3$ yield $\delta = 0.5$.
Note that in the case of Fig.\ref{fig1} $\mu=5$ and $\mu'=2.41$.  
The breakdown of Eq.(\ref{nomemory}) is referred to by us 
as \emph{memory beyond memory} effect. In fact, 
the existence of pseudo events  implies correlation among different 
times of the series $\{\tau_{i}\}$, and thus a memory of earlier events. 
This memory is annihilated by shuffling the order of these times, as 
we do with a bunch of cards. In fact, as shown by the inset 
of Fig.\ref{fig1}, we see that shuffling has the effect of yielding 
$\delta = 0.5$, in 
accordance with the theoretical prescription of Eq.(\ref{nomemory}). 
It is impressive that the scaling detected by the DE method does not depend 
on the pseudo events, but only on the hidden events that would be 
invisible to an analysis based on the direct evaluation of $\psi(\tau)$.

Let us apply this model to the real data taken from \cite{heartbeatingdata}.
We apply our technique to 33 long-time ECG records (about 20 hours each), 
18 healthy and 15 
showing pathological behavior related to Congestive Heart Failure (c.h.f.). 
Following Ref.\cite{heartbeatingdata}, we refer to all the ECG records of the
{\em MIT-BIH Normal Sinus Rhythm Database} and of the
{\em BIDMC Congestive Heart Failure Database}, for the 
healthy and the c.h.f. patients, respectively.

\begin{figure}[h]
\begin{center}
\includegraphics[angle=270, scale=.31]{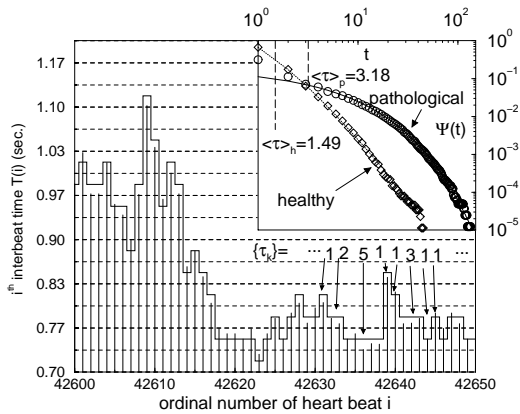}
\caption{\label{fig2} The inter-beat time 
$T(i)$ as a function of the
number of beats, $i$. The thick line denotes the trajectory
corresponding to the coarse graining given by $\Delta T = 1/30$ sec. The
vertical lines denote the height of the original data. 
The arrows and the integer labels illustrates 
how the sequence of $\tau_{k}$'s is generated. 
Inset: Survival probabilities. The circles denote
the c.h.f. patients and the corresponding fitting function is 
$\Psi(t) = 0.19 \exp[-(t/3.1)^{0.6}]$. 
The diamonds denote the healthy patient
and the corresponding fitting function is
$\Psi(t)=5.71/(0.93+t)^{3.25}$.}
\end{center}
\end{figure}

The data under study are time 
series of the kind of that illustrated in Fig. 2, where the 
length of the vertical lines expresses 
$T(i) = t_{i} - t_{i-1}$ as a function of the integer number $i$. 
The integer number $i$ denotes the i-th heart beating of an 
electrocardiogram, and $t_{i}$ is the time at which the R wave of this 
heart beat occurs.
\vspace{.3 cm}
	 We make these data suitable for the 
         illustration of the \emph{memory beyond memory} effect as 
         follows. We adopt a procedure illustrated with the help of 
         Fig.\ref{fig2}. The vertical axis, concerning the variable $T(i)$, is 
         divided into many cells of a given size $\Delta T$. Thus the 
         $(T(i),i)$-plane is divided into many horizontal strips with a 
         constant width equal to $\Delta T$. This coarse-graining 
         prescription yields the thick line of Fig.\ref{fig2}. The curve 
         corresponds to many horizontal intervals separated by 
         vertical up and down jumps.
The widths of these horizontal intervals define a sequence of numbers 
$\tau_{i}$ that is the object of our statistical analysis. To make this 
analysis as efficient as possible we have to make a proper choice of 
the value of $\Delta T$, since an excessively small value would produce 
too many pseudo events and an excessively large  would yield poor 
statistics. The results of our statistical analysis were proven to be 
insensitive to changing $\Delta T$ over a wide range of values. 
We therefore assign to $\Delta T$ the mean value of this range, 
which turns out to be $\Delta T = 1/30$ sec.

\begin{figure}[h]
\begin{center}
a)\includegraphics[angle=270, scale=.31]{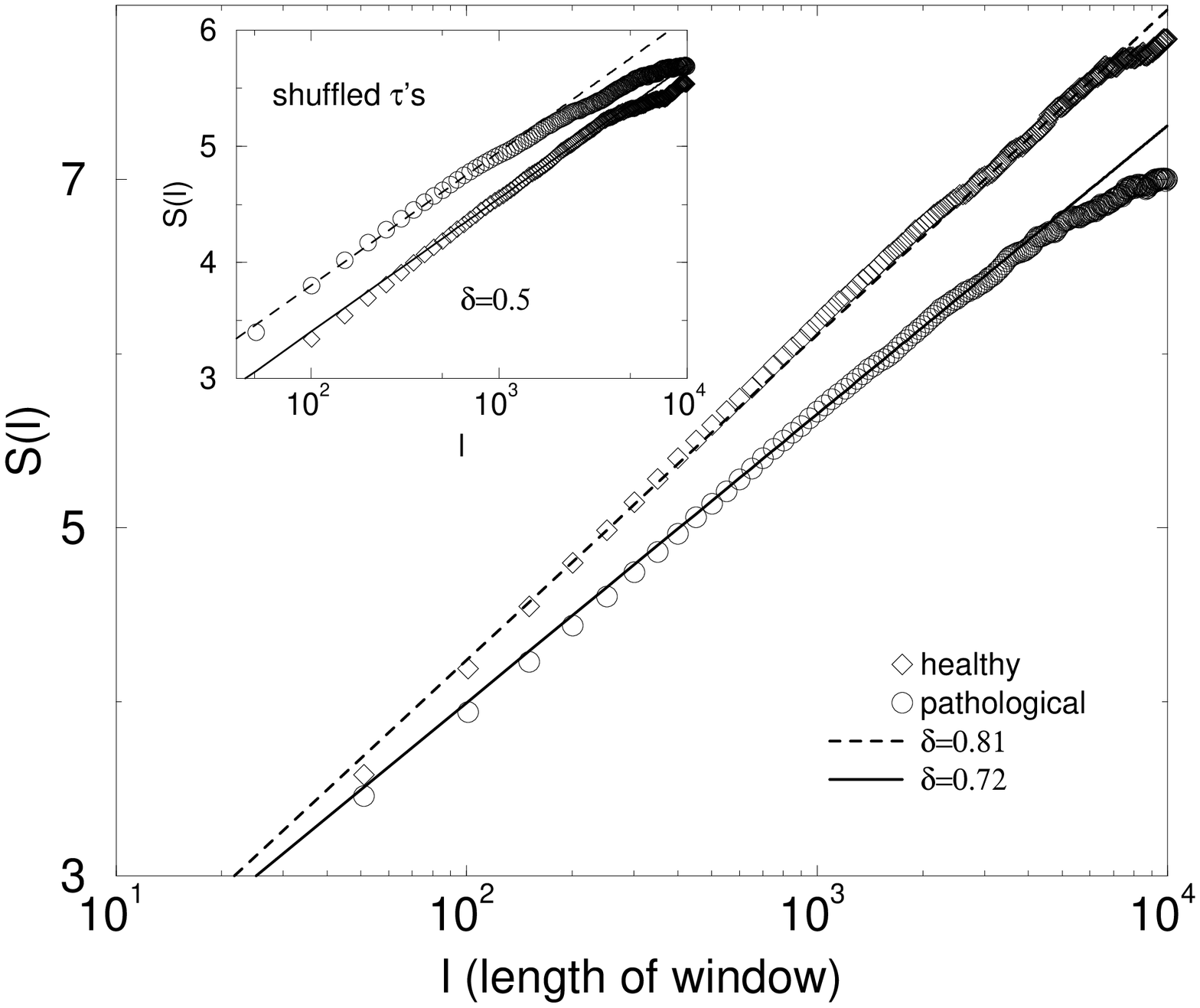}
b)\includegraphics[angle=270, scale=.30]{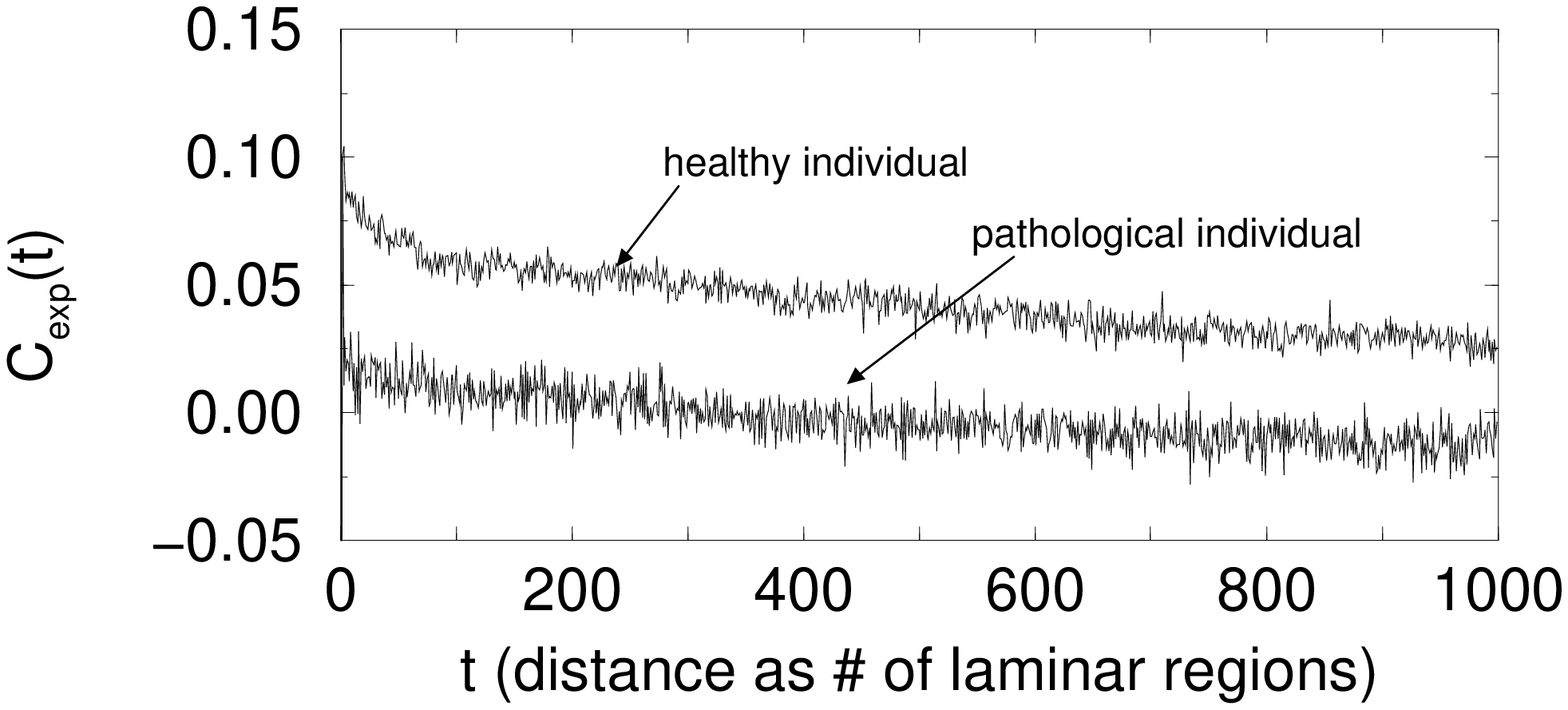}
\caption{\label{figura3}
a) The DE as a function of $l$. The inset illustrates the effect 
of shuffling, the two lines correspond to $\delta= 0.5$.
b) The correlation function $C_{exp}(t)$ as a function of $t$ on a
healthy and on a c.h.f. individual. }
\end{center}
\end{figure}

         The events under study refer to the jumps from one to another 
strip. To assess whether they are events or pseudo events, we have to
compare the waiting time distribution $\psi(\tau)$ to the scaling 
detected by means of the DE method.
For the sake of statistical accuracy we decided to evaluate the 
probability of finding waiting times larger than a given value $\tau$. 
This is the function $\Psi(\tau)$ defined by 
$\Psi(\tau) \equiv \int _{t}^{\infty} {\rm d}\tau \psi(\tau)$. 
The results illustrated in the 
inset of Fig.\ref{fig2} imply the Brownian scaling $\delta = 0.5$. In fact, the 
function $\Psi(\tau)$ of the heart failure subject is a stretched 
exponential and the healthy subject yields $\mu = 3.9$. The same Brownian
condition applies to all the subjects. On the contrary, the DE method 
yields for the healthy subjects the mean value $\delta = 0.82 \pm 0.04$ 
and for the heart failure subject $\delta = 0.71 \pm 0.06$. It is 
interesting to notice that Fig.\ref{figura3} refers to the same subjects as those 
of the inset of Fig.\ref{fig2}, and yields for the healthy $\mu'= 2.17$ and for 
the heart failure subject $\mu' = 2.4$. If we shuffle the numbers of the 
sequence {$\tau_{i}$} we recover $\delta = 0.5$, a fact proving that the 
\emph{memory beyond memory} effect is a genuine property of heartbeat.
	
The additional memory is confirmed by the numerical evaluation 
of the normalized correlation function of the variable $\tau_i -\langle \tau_i \rangle$, 
denoted by $C_{exp}(t)$, where the symbol $t$ is the continuous approximation 
of the discrete patch label $i$. The two-variables model that we are 
using to explain the \emph{memory beyond memory} effect would yield
	 \begin{equation}
	     \label{correlationfunction}
	     C(t) \propto 1/t^{\beta}.  
	     \end{equation}
The index $\beta$ in that case would be a complicated function of the 
four parameters involved by the two-variable model. This means that 
the DE is a memory detector more efficient and much less ambiguous 
than the correlation function. 
The DE selects  from the distribution of times described by the 
arbitrary $\psi(\tau)$ the really random events yielding a non 
arbitrary distribution with a  unique $\mu'$.
The correlation function $C(t)$, 
on the contrary, depends on the details of the model, but does not 
afford an easy way to define them. For the main purpose of this letter
it is enough to point out
that the form of the correlation function $C_{exp}(t)$ is
	    \begin{equation}
		 C_{exp}(t) = (1-\epsilon^2) W(t) +\epsilon^2 C(t).
		 \label{ratherthancmm}
		 \end{equation} 		
Here $W(t)$ denotes a function dropping from $1$ to $0$ in one time step, 
while the function $C(t)$, with the asymptotic form of Eq.(\ref{correlationfunction}), 
is continuous for for $t \rightarrow 0$.

We account for the structure of Eq.(\ref{ratherthancmm}) as follows. 
The sequence $\{\tau_{i}\}$ is generated by the joint use of two models. The first is 
the model of Eq.(\ref{keyequation}) with no additional variables and no 
{\em memory beyond memory} property, the 
second is the model with two variables.
These two models generate two independent 
sequences $\{\tau_{i}\}$. 
To any index $i$ we assign, with probability $\epsilon$,
 the value provided by the model with additional memory, and,
with probability $1-\epsilon$, the value provided by 
the model with only one variable.
This model is 
reminiscent of one adopted to account for the statistical 
properties of DNA sequences \cite{maria}.
The function $C_{exp}(t)$ in one step drops from the value $C_{exp}(0) = 1$ 
to the value $C_{exp}(1) = \epsilon^{2} C(1) \simeq \epsilon^2$, thereby allowing us 
to derive $\epsilon$ from the experimental 
correlation function at $t = 1$.

\begin{figure}[h]
\begin{center}
\includegraphics[angle=270, scale=.31]{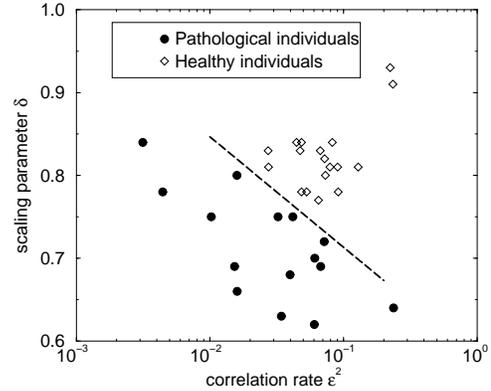}
\caption{\label{fig5}
Values of the scaling parameter $\delta$ and 
of $\epsilon^2$ 
for the healthy and c.h.f. individuals
of this analysis.} 
\end{center}
\end{figure}

In conclusion, the meaning of the parameter $\epsilon$ is as 
follows. The value $\epsilon = 1$ would imply that the heart beating is 
described only by a model with two variables, $x$ and $y$.
In other words, the larger $\epsilon$ the larger the weight 
of the \emph{memory beyond memory} effect. The parameter $\delta$ is
connected to the time distance between two nearest-neighbor real 
events.
If this time 
distribution is exponential, there is no memory of conventional type, 
as earlier observed. If $\mu'$ becomes closer and closer to $\mu' = 2$, 
this conventional memory becomes stronger and stronger. Thus, to 
establish a more intuitive understanding of what happens to memory, 
regardless of whether it is of the conventional or of the new type, 
let us adopt the following perspective. The condition of highest 
memory  corresponds to $\epsilon = 1$ and $\delta = 1$. This would mean 
that the heart beating depends only on the {\em memory beyond 
memory} model, and, that, at the same time, $\mu' 
= 2$. The opposite case, of complete absence of memory, implies 
$\epsilon = 0$ and  $\mu' \rightarrow \infty$. This would mean that the heart 
beating is very well modeled by the one-variable  model of Eq.(\ref{keyequation}), 
with an exponential distribution of waiting times, in other words, 
without any memory of whatsoever form. This leads us to express
the distribution of  patients in the ($\delta$, $\epsilon^2$)-plan of Fig.\ref{fig5}.
We note the surprising result that all 
the healthy subjects and all the heart failure subjects are contained 
in the top-right region and
in the bottom-left one, respectively. We also notice that  all 
the healthy subjects, but two of them, are localized within a small 
portion of the top-right region of the graph, not far from the border with the 
heart failure region. We have the impression that this reflects the 
fact that the healthy function of the heart beating system depends 
on a proper balance of memory and randomness that the analysis of 
this letter makes ostensible.
The distribution of the heart failure subjects within the bottom-left 
region is much broader. It would be desirable to have at our 
disposal the patient survival probability as, for
example, in Ref. \cite{danish},
to assess whether a physiological reaction to the c.h.f. pathology, responsible
for the bottom-left broadened distribution, plays a negative or a 
positive role.
 The advocates of the second possibility might argue 
that higher randomness and broader distribution reflect an 
effort of the perturbed heart beating system to explore all possible 
states to recover the lost function.

        It is difficult to establish a connection with the earlier 
research work in this field since, to the best of our knowledge, the 
\emph{memory beyond memory} effect was never observed, and we did not 
find any explicit mention to it in the field of heart beating. 
Is there a correspondence between the {\em memory beyond memory} effect and 
the multifractal properties observed in Ref.\cite{Stanleynature}? 
We are inclined to 
believe that there is, due to the similarity between Fig.\ref{fig5}
and Fig.4b of \cite{Stanleynature}.

As to the memory property expressed by $\delta$, and by the 
corresponding $\mu'$, as well, we would like to mention Ashkenazy {\em et al.} 
\cite{stanleybemerciful} whose results show for the healthy patients
a deviation from ordinary scaling higher than that of c.h.f patients.
If we identify the scaling detected by these authors in the case of 
magnitude fluctuations and in the large-time regime, with the scaling 
parameter $\delta$ determined by the DE method, we see that our $\delta
= 0.82 \pm 0.04$ for the healthy subjects corresponds to their $\delta = 0.82$, 
and that our $\delta = 0.71 \pm 0.06$ for the heart failure subjects corresponds 
to their $\delta = 0.71$. Thus, we conclude that our findings do not 
conflict, or do not necessarily do, with the earlier findings.


\begin{references}
    \bibitem{shlesinger} M. F. Shlesinger, Ann. N. Y. Acad. SCi. 504, 214 (1987); 
J. B. Bassingthwaighte, L. S. Liebovitch, and B. J. West, 
\emph{Fractal Physiology} (Oxford University Press, New York, 1994).
    \bibitem{nicola}
    N. Scafetta, P. Hamilton, P. Grigolini,  Fractals, {\bf 9}, 193 (2001).
    \bibitem{giacomo}
P. Grigolini, L. Palatella, G. Raffaelli, in press on Fractals 
(2001); cond-mat/0104166.
\bibitem{vito} N. Scafetta, V. Latora, P. Grigolini, cond-mat/0105041.
\bibitem{katia} D. Bedeaux, K. Lakatos Lindenberg and K. E. Shuler, J. 
Math. Phys. {\bf 12}, 2116 (1971). 
\bibitem{heartbeatingdata}
 A. L. Goldberger, et al.
{\em PhysioBank, PhysioToolkit, and Physionet: Components of a New Research Resource} 
{\em for Complex Physiologic Signals.}
Circulation {\bf 101(23)}: e215-e220 
(Circulation Electronic Pages; http://circ.ahajournals.org/cgi/content/full/101/23/e215); 
2000 (June 13). 
\bibitem{maria}
P. Allegrini, M. Buiatti, P. Grigolini and B.J. West,
Phys. Rev. E {\bf 57}, 4558 (1998).
\bibitem{danish} K. Saermark, M. Moeller, U. Hintze, H. Moelgaard, P. E. Bloch 
Thomsen, H. Huikuri, T. Makikiallio, J. Levitan and M. Lewkowicz, 
Fractals {\bf 8}, 315 (2000). 
\bibitem{Stanleynature} P. Ch. Ivanov, L. A. Nunes Amaral, A. L. Goldberger, S. Havlin, M. G. 
Rosenblum, Z. R. Struzik and H. E. Stanley, Nature {\bf 399}, 461 
(1999).
\bibitem{stanleybemerciful} Y. Ashkenazy, P. Ch. Ivanov, S. Havlin, 
C. -K. Peng, A. L. Goldberger, and H. E. Stanley, Phys. Rev. Lett. {\bf 
86}, 1900 (2001). 
\end{references}
\end{document}